\documentclass[12pt]{article}  
\usepackage{amsmath}
\usepackage{graphicx}
\usepackage{natbib}
\usepackage{float}
\usepackage{subcaption}
\usepackage{authblk}
\usepackage{booktabs}
\usepackage[none]{hyphenat}
\usepackage{setspace} 
\usepackage[paperwidth=8.7in, paperheight=11in,margin=0.8in]{geometry}
\usepackage[hidelinks]{hyperref}
\title{\textbf{Integrative Prognostic Modeling of Breast Cancer Survival with Gene Expression and Clinical Data}}

\author[1]{Robert Amevor*}
\author[2]{Emmanuel Kubuafor}
\author[2]{Dennis Baidoo}
\author[4]{Junaidu Salifu}
\author[2]{Koshali Muthunama Gonnage}
\author[3]{Onyedikachi Joshua Okeke}

\affil[1]{Arnold School of Public Health, University of South Carolina, Columbia, SC, USA}
\affil[2]{Department of Mathematics and Statistics, University of New Mexico, Albuquerque, NM, USA}
\affil[3]{Department of Geography and Environmental Studies, Texas State University, San Marcos, TX, USA}
\affil[4]{College of Population Health, University of New Mexico, Albuquerque, NM, USA}

\date{} 

\begin{document}

\maketitle

\noindent\textbf{*Corresponding author:} amevorrobert80@gmail.com

\section*{Abstract}

\textbf{Background:}Accurate survival prediction in breast cancer is essential for patient stratification and personalized therapy. Integrating gene expression data with clinical factors may enhance prognostic performance and support precision medicine.

\textbf{Objective:}To develop an integrative survival prediction model combining clinical variables and gene expression signatures in breast cancer, and to assess the relative contribution of each component using penalized Cox regression and machine learning. \textbf{Methods:}We analyzed 1,867 patients from the METABRIC cohort with clinical annotations and microarray-based gene expression profiles. The top 5,000 most variable genes (median absolute deviation) were retained. Elastic Net–penalized Cox regression identified 75 predictors (70 genes and 5 clinical variables: tumor size, stage, surgery type, age at diagnosis, and Nottingham Prognostic Index). Model performance was evaluated with Harrell’s concordance index (C-index) and 36-month time-dependent AUC. Random Survival Forests (RSF) were trained on the top 20 genes to assess nonlinear effects and validate variable importance. Gene expression patterns were visualized using PCA and heatmaps. \textbf{Results:}The integrative Cox model achieved a C-index of 0.922 and a 36-month AUC of 0.94, outperforming clinical-only (C = 0.64). RSF analysis confirmed prognostic contributions of top genes (e.g., OR2T27, TBATA, LINC01165, SLC10A4), yielding a 36-month AUC of 0.88. PCA and heatmaps revealed distinct molecular profiles across risk groups. \textbf{Conclusions:}Integrating gene expression signatures with clinical variables substantially improves survival prediction in breast cancer. This framework offers a robust foundation for individualized prognostic assessment and may inform clinical decision-making and trial design.

\textbf{Keywords}:Breast cancer, gene expression, Elastic Net, Cox model, Random Survival Forests, integrative modeling, survival prediction.

\section{Introduction}
Breast cancer remains the most commonly diagnosed cancer among women worldwide, accounting for over 2.3 million new cases and 685,000 deaths annually as of 2020 \cite{Siegel2023, Sung2021}. Despite advances in early detection and systemic therapies, survival outcomes remain heterogeneous even among patients within the same clinical stage or molecular subtype. This heterogeneity underscores the urgent need for more accurate and individualized prognostic tools that go beyond traditional staging to better inform treatment decisions and patient counseling.

Historically, prognostic models for breast cancer have relied on clinicopathologic variables such as tumor size, nodal status, histological grade, and hormone receptor expression \cite{Goldhirsch2013}. While these factors provide valuable information, they often fall short in accounting for the biological complexity of breast tumors. High-throughput technologies, particularly gene expression profiling, have significantly advanced our understanding of tumor biology. The seminal work by \cite{Perou2000} introduced the concept of molecular subtypes in breast cancer, demonstrating that tumors of similar histology can differ profoundly at the genomic level .

Building on this molecular foundation, commercial assays such as Oncotype DX and MammaPrint have been clinically validated to assess recurrence risk and guide chemotherapy use in early-stage breast cancer \cite{Sparano2018, Paik2004}. However, these tools are limited by fixed gene panels, high costs, and potential lack of generalizability across diverse populations. Consequently, there has been a growing interest in developing integrative prognostic models that incorporate clinical, genomic, and immune-related variables using advanced machine learning techniques   \cite{Varnier2021,Hussain2024,Sammut2022}.

Elastic Net penalized Cox regression is particularly suitable for high-dimensional omics data as it performs variable selection while handling multicollinearity among gene features \cite{Zou2005}. Additionally, Random Survival Forests (RSF) offer a non-parametric alternative that can capture complex nonlinear interactions without stringent modeling assumptions \cite{Ishwaran2008, Boulesteix2012}. These approaches have been increasingly used in cancer prognostic modeling and have demonstrated robust performance in both internal and external validation settings \cite{Wang2024,Hao2021}.

In this study, we aimed to develop an integrative prognostic model for breast cancer survival by combining clinical features with gene expression signatures, including immune-related markers. We applied Elastic Net Cox regression for gene selection and RSF for non-linear modeling and variable importance assessment. Visualization methods, including heatmaps and principal component analysis (PCA), were used to explore molecular patterns associated with risk stratification. Our goal was to evaluate whether this integrative modeling framework could outperform models using clinical or gene-only inputs, thereby supporting more personalized and biologically informed patient care.

\section{Methods}

\subsection{Data Sources}
We utilized a publicly available cohort (BRCA METABRIC) comprising approximately 1,979 breast cancer patients with matched clinical data and gene expression profiles derived from microarray technology. The dataset included relevant clinical variables such as lymph node status, tumor size, type of breast surgery, and overall survival time and status. Gene expression values were log-transformed and standardized across samples.

\subsection{Data Preprocessing}
Clinical covariates with missing values were imputed using multiple imputation by chained equations (MICE), implemented via the \texttt{mice} R package. Each incomplete variable was imputed using a fully conditional specification appropriate to its type (e.g., predictive mean matching for continuous variables and logistic regression for binary variables). Five imputed datasets were generated, and results were pooled according to Rubin's rules.

To reduce noise and computational burden, genes with low variance across samples were excluded. The top 5,000 most variable genes—based on median absolute deviation (MAD)—were retained for downstream modeling, a commonly used threshold balancing dimensionality reduction and biological signal retention.

\subsection{Feature Selection with Elastic Net Penalized Cox Model}
To identify a parsimonious subset of prognostic genes, we applied the Elastic Net regularization method within a Cox proportional hazards framework. The Elastic Net combines both Lasso ($\ell_1$) and Ridge ($\ell_2$) penalties, promoting sparsity while handling multicollinearity among predictors. The optimization objective is given by:

\[
\hat{\beta} = \arg\min_{\beta} \left\{ - \ell(\beta) + \lambda \left( \alpha \lVert \beta \rVert_1 + \frac{1 - \alpha}{2} \lVert \beta \rVert_2^2 \right) \right\}
\]

where $\ell(\beta)$ is the partial log-likelihood of the Cox model, $\lambda$ is the regularization parameter controlling overall penalty strength, and $\alpha$ balances the contribution of the Lasso and Ridge penalties.

To balance the benefits of L1 (Lasso) and L2 (Ridge) regularization, we set the Elastic Net mixing parameter $\alpha = 0.5$ using the \texttt{glmnet} R package, giving equal weight to both penalties. This value was selected to promote both sparsity (variable selection) and grouping effects among correlated predictors, which are common in gene expression data. While values of $\alpha$ ranging from 0.1 to 0.9 were explored in preliminary analysis, the overall model performance—measured by cross-validated partial likelihood deviance and stability of selected features—was comparable across this range. Hence, $\alpha = 0.5$ was adopted as a pragmatic default to balance model interpretability and robustness, consistent with previous integrative modeling studies in high-dimensional omics data.

 A sequence of candidate $\lambda$ values was generated on a log scale, and the optimal value was selected via 10-fold cross-validation based on the partial likelihood deviance. This approach ensured that the model selected a robust and interpretable set of gene predictors with optimal predictive performance.

Model performance was further evaluated using the time-dependent concordance index (C-index) and time-dependent area under the curve (AUC) at clinically relevant time points (12, 24, and 36 months), computed across folds. The final model was refit using the optimal $\lambda$ on the full dataset.

\subsection{Multivariable Cox Proportional Hazards Model}
The genes selected by Elastic Net were then entered together with key clinical variables in a multivariable Cox proportional hazards model. The model has the form:

\[
h(t \mid X) = h_0(t) \exp\left( \beta_1 X_1 + \beta_2 X_2 + \dots + \beta_p X_p \right)
\]

where $h(t \mid X)$ is the hazard at time $t$ given covariates $X$, and $h_0(t)$ is the baseline hazard. Hazard ratios and 95\% confidence intervals were reported for all selected predictors.

\subsection{Nonlinear Validation with Random Survival Forests}
To validate and assess nonlinear relationships among variables, a Random Survival Forest (RSF) was trained on the same data. RSF grows an ensemble of survival trees using bootstrapped samples, with each split optimizing the log-rank test statistic. Variable importance measures were computed using the permutation-based approach, defined as:

\[
V_j = \frac{1}{B} \sum_{b=1}^{B} \left( \widehat{C}^{(b)} - \widehat{C}^{(b,j)} \right)
\]

where $\widehat{C}^{(b)}$ is the concordance of tree $b$ with the original data and $\widehat{C}^{(b,j)}$ is the concordance after permuting feature $j$.

\subsection{Model Performance Evaluation}
Model performance was evaluated using 10-fold cross-validation, repeated 3 times to ensure stability. For each fold, the training set was used to fit the model and the test set was used to estimate the concordance index (C-index) and time-dependent receiver operating characteristic AUC at clinically relevant time points (12, 24, and 36 months). The C-index measures the proportion of concordant pairs among all comparable pairs of subjects, with values closer to 1 indicating better discrimination. We report average metrics across folds. To further validate robustness, we performed bootstrap resampling (B = 100) on the validation cohort to estimate optimism-corrected performance.

\subsection{Visualization}
To explore patterns in the multi-omics data and assess group separation, we performed principal component analysis (PCA) on the standardized gene expression matrix. PCA reduces dimensionality by projecting data onto orthogonal directions that maximize variance. Formally, the principal components are obtained by solving:
\[
\max_{w} \quad w^\top \Sigma w \quad \text{subject to} \quad \|w\| = 1
\]
where $\Sigma$ is the sample covariance matrix of the standardized gene expression data.

The first two principal components were plotted to visualize clustering patterns among patients stratified by predicted risk groups from the Cox model.

In addition, we generated heatmaps of the top selected genes (e.g., P2RX2, SOCS6, OR2T27, GPR31, ATOH7, RXFP2) using hierarchical clustering. The pairwise similarity between patients was measured using the Euclidean distance:
\[
d(x_i, x_j) = \sqrt{ \sum_{k=1}^{p} (x_{ik} - x_{jk})^2 }
\]
where $x_i$ and $x_j$ are standardized gene expression vectors for patients $i$ and $j$.

Clusters were then merged using complete linkage, which defines the distance between two clusters $A$ and $B$ as:
\[
D(A,B) = \max_{x \in A, \, y \in B} d(x,y)
\]
to form more compact, spherical clusters.  

Patients were ordered by predicted risk score, and gene expression levels were scaled by z-score normalization:
\[
z_{ik} = \frac{x_{ik} - \mu_k}{\sigma_k}
\]
with $\mu_k$ and $\sigma_k$ denoting the mean and standard deviation of gene $k$, to emphasize relative differences.  

All visualizations aimed to provide an interpretable view of how high- and low-risk patients differed at the molecular level, supporting the results of the survival models.

\section{Results}

\subsection{Study Population}
A total of 1900 breast cancer patients with available clinical and microarray gene expression data were included in the analysis. Following quality control and removal of samples with missing or incomplete data, 1867 patients were retained for modeling. The median age at diagnosis was  61.80 years (range from 22-97), with a distribution across tumor stages I to IV. Patients underwent a range of surgical interventions, including breast-conserving surgery and mastectomy, and received various adjuvant treatments. Key clinical variables analyzed included tumor size, lymph node status, type of breast surgery, and relapse-free survival time.

\subsection{Feature Selection via Elastic Net Cox Regression}
To identify a parsimonious yet robust set of prognostic biomarkers, we applied Elastic Net penalized Cox regression incorporating both clinical variables and gene expression data. Cross-validation was performed to select the optimal penalty parameter, minimizing prediction error.

The Elastic Net cox model retained approximately 85 features at the optimal lambda (Figure 1), including a focused subset of around 20 genes with the strongest prognostic contributions. Notable selected genes included OR2T27, KLRC3, TBATA, LINC01165, SOCS6, P2RX2, GPR31, MBD3L2, and ATOH7. While some of these genes are not yet widely reported in breast cancer, they have emerging links to immune signaling, receptor activity, and transcriptional regulation, suggesting potential roles in tumor biology that merit further investigation. Additionally, key clinical predictors such as Nottingham prognostic index, tumor stage, age at diagnosis, and lymph node positivity remained highly informative.

\begin{figure}[H]
\centering
\includegraphics[width=0.8\textwidth]{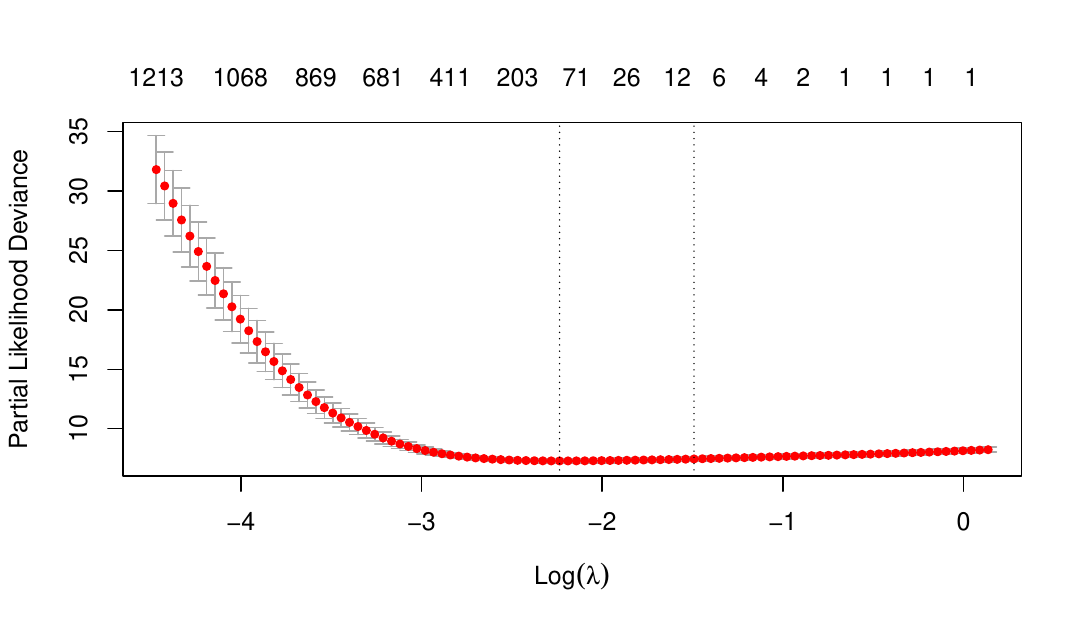}
\caption{10-fold cross-validation curve for Elastic Net Cox regression. The optimal value of the regularization parameter $\lambda$ was selected at the minimum mean cross-validated partial likelihood deviance.}
\label{partial_likelihood_deviance}
\end{figure}

The risk scores generated by the Elastic Net Cox model stratified patients into tertiles—low, medium, and high risk—with clear separation in survival outcomes. This stratification was validated using Kaplan–Meier survival curves ( Figure 2)  and demonstrated strong predictive performance, with a concordance index (C-index) of 0.922 and time-dependent area under the curve (AUC) values of 0.947 at 12 months, 0.951 at 24 months, and 0.949 at 36 months.

\begin{figure}[H]
\centering
\begin{subfigure}{0.48\textwidth}
    \centering
    \includegraphics[width=\linewidth]{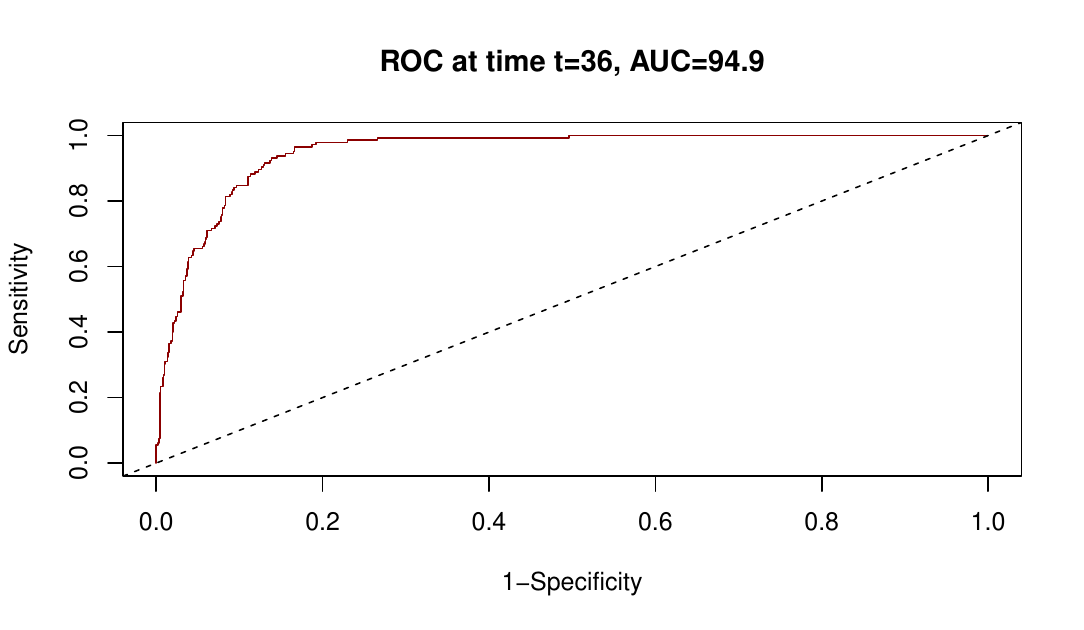}
    \caption{Time-dependent AUC curves showing the performance of the integrative Cox model over 36 months. The model achieved an AUC of 0.949 at 36 months, indicating high discriminative ability.}
    \label{fig:auc_curve}
\end{subfigure}
\hfill
\begin{subfigure}{0.48\textwidth}
    \centering
    \includegraphics[width=\linewidth]{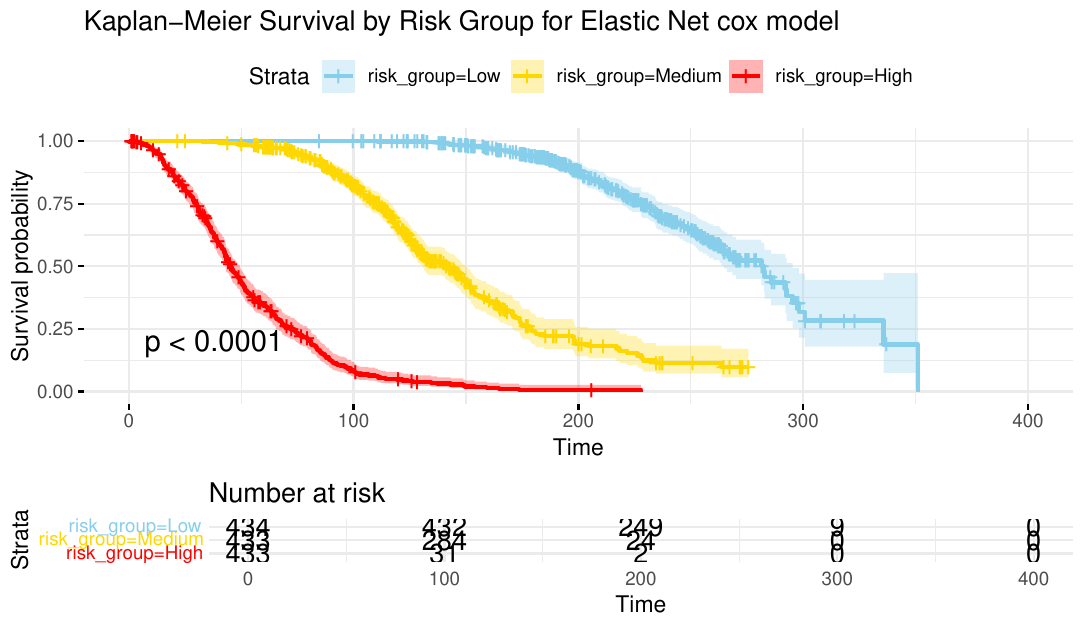}
    \caption{Kaplan–Meier survival curves for breast cancer patients stratified into low, medium, and high-risk groups based on Elastic Net-derived risk scores. Clear separation confirms effective risk discrimination.}
    \label{fig:km_curve}
\end{subfigure}
\caption{Comparative evaluation of the integrative Cox model. (a) Time-dependent AUC demonstrating high discriminative ability. (b) Kaplan–Meier curves confirming effective risk stratification.}
\label{fig:combined_figures}
\end{figure}

\subsection{Random Survival Forest (RSF) Results}
To evaluate nonlinear relationships and validate prognostic markers, the Random Survival Forest (RSF) model was trained using 500 trees, each grown with a resampling size of 822 patients without replacement. The terminal node size was set to 15, and five variables were considered at each split. The RSF achieved an out-of-bag (OOB) error rate of approximately 33.7\%, indicating reasonably good discrimination. The model’s standardized Continuous Ranked Probability Score (CRPS) was 0.163, suggesting adequate calibration of survival probabilities across time.

Variable importance analysis ( Figure 3) identified both clinical and molecular features as key predictors. The most important variables included Nottingham prognostic index (0.127), tumor stage (0.089), and gene expression values for OR2T27, TBATA, and LINC01165, all of which exhibited variable importance scores exceeding 0.04.

\begin{figure}[H]
\centering
\includegraphics[width=0.8\textwidth]{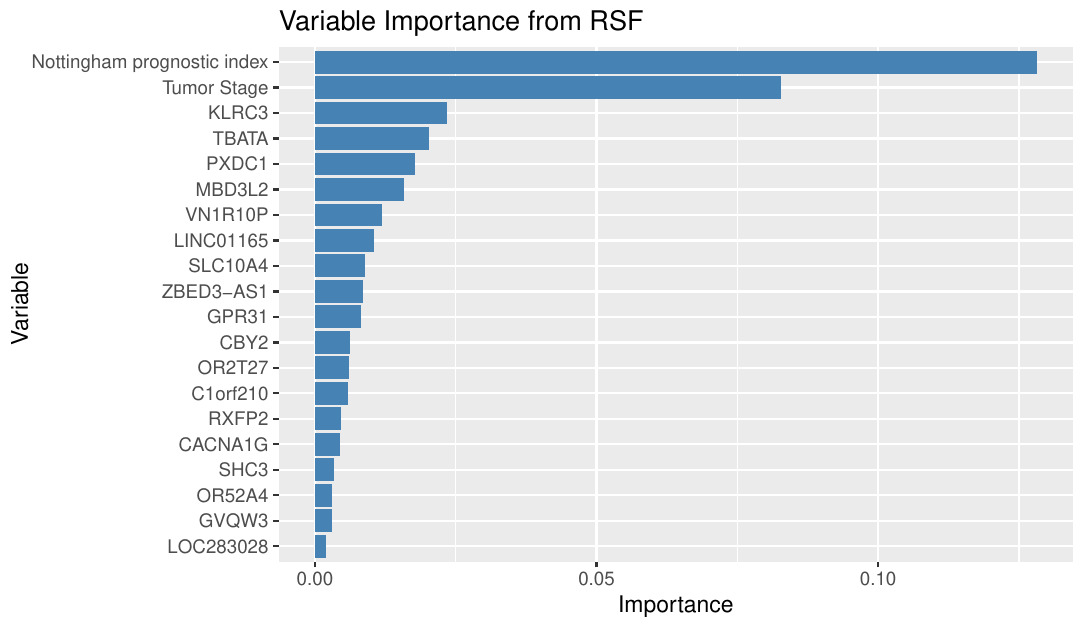}
\caption{Top 20 predictors identified by Random Survival Forests based on variable importance scores. Both clinical and genomic features contribute to survival prediction.}
\label{Variable_Importance_plot}
\end{figure}

To further evaluate model performance, we estimated RSF-predicted median survival times across risk groups. The predicted median survival was 295.3 months for the low-risk group, 153.9 months for the medium-risk group, and 51.7 months for the high-risk group. Kaplan-Meier curves stratified by RSF-derived risk groups also showed significant separation (log-rank p $<$ 0.001) with median predicted survival times.  This clear stratification further supports the clinical utility of the integrated prognostic model.
Time-dependent ROC analysis (Figure 4) yielded an AUC of 0.88 at 36 months, indicating strong discriminative ability, though slightly lower than the Elastic Net Cox model (AUC = 0.949).

\begin{figure}[H]
\centering
\includegraphics[width=0.8\textwidth]{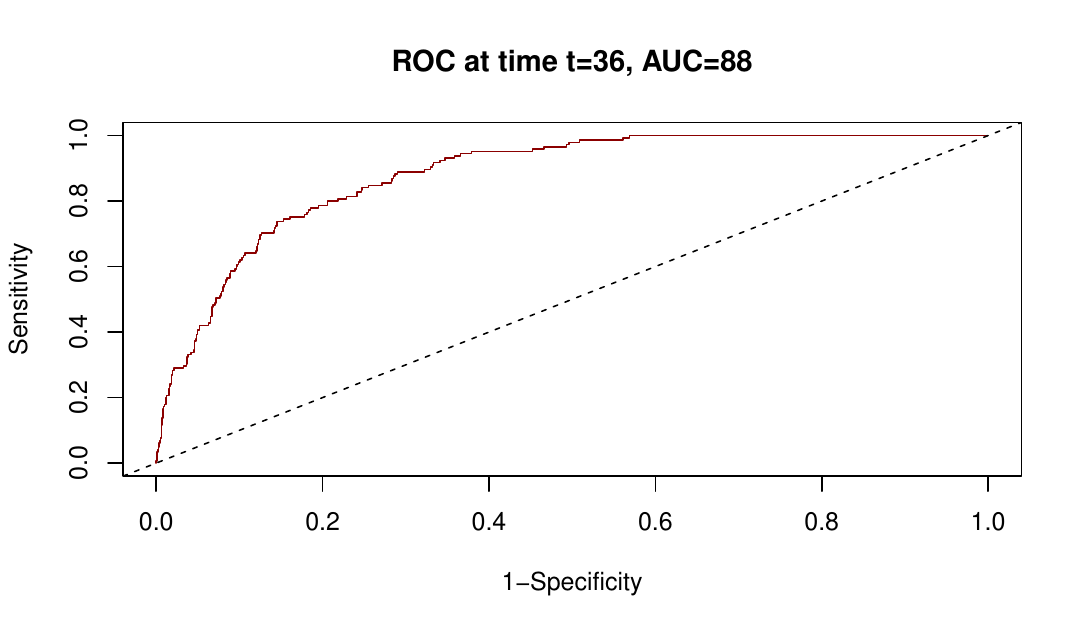}
\caption{Time-dependent AUC curves showing the performance of the Random Survial Forest over 36 months.The model achieved an AUC of 0.88 at 36 months,indicating high discriminative ability.}
\label{geneplot}
\end{figure}

\subsection{Multivariable Cox Proportional Hazards Model Results and Sensitivity Analyses}
To confirm the robustness of the prognostic findings, we refitted multivariable Cox models incorporating the strongest gene predictors along with relevant clinical variables

First, a parsimonious multivariable Cox model included the key clinical features (lymph node positivity, tumor size, type of breast surgery) along with high-ranking genes (P2RX2, SOCS6, OR2T27, GPR31, ATOH7, RXFP2) identified through Elastic Net. This model demonstrated a concordance index (C-index) of 0.641 (SE = 0.009), supporting moderate discriminative ability. Several gene features such as SOCS6, OR2T27, and GPR31 retained statistical significance within this model, indicating their added prognostic value beyond standard clinical variables.  The hazard ratio for P2RX2 was estimated at 1.1328, indicating its role as an adverse prognostic marker. Similarly, SOCS6 and OR2T27 displayed statistically significant hazard ratios of 1.7255 and 1.5733, respectively, with GPR31 having a harzard ratio 0.4764 suggesting their relevance to breast cancer survival.

A forest plot (figure 5) of the multivariable Cox model visually summarized the effect sizes and confidence intervals of key predictors, showing strong protective and risk-enhancing contributions from several genes.  Likelihood ratio, Wald, and score tests were all highly significant (p $<$ 0.001), suggesting strong overall model fit.

\begin{figure}[H]
\centering
\includegraphics[width=0.7\textwidth]{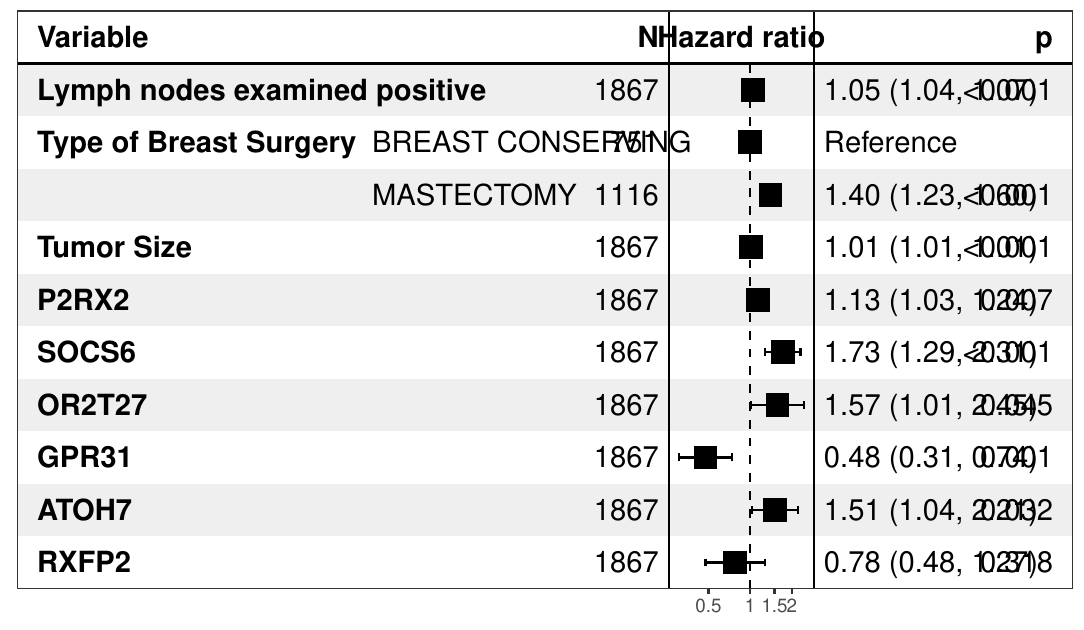}
\caption{Forest plot from the reduce multivariable Cox proportional hazards model showing hazard ratios (HRs) and 95\% confidence intervals for selected clinical and gene expression predictors of breast cancer survival}
\label{coxfinal_forest}
\end{figure}

Proportional hazards assumptions were checked using Schoenfeld residuals, with no major violations detected globally (p = 0.31), though a few covariates showed minor deviations. Additionally, flexible modeling with natural cubic splines on tumor size suggested non-linear effects, slightly improving the model concordance to 0.642 (SE = 0.009) while maintaining similar variable significance patterns.

As a sensitivity analysis, a reduced Cox model including only the most clinically actionable variables (type of breast surgery, tumor size, P2RX2, SOCS6, OR2T27, GPR31, ATOH7) was fitted, yielding a C-index of 0.61, confirming that simplified models could still provide meaningful stratification while being more interpretable in clinical settings.

Furthermore, time-dependent AUC analysis (Figure 6) for the reduced cox proportional model showed consistent discrimination at multiple follow-up horizons: 12-month AUC 0.652, 24-month AUC 0.650, and 36-month AUC 0.681, demonstrating temporal stability of the prognostic model.

\begin{figure}[H]
\centering
\includegraphics[width=0.65\textwidth]{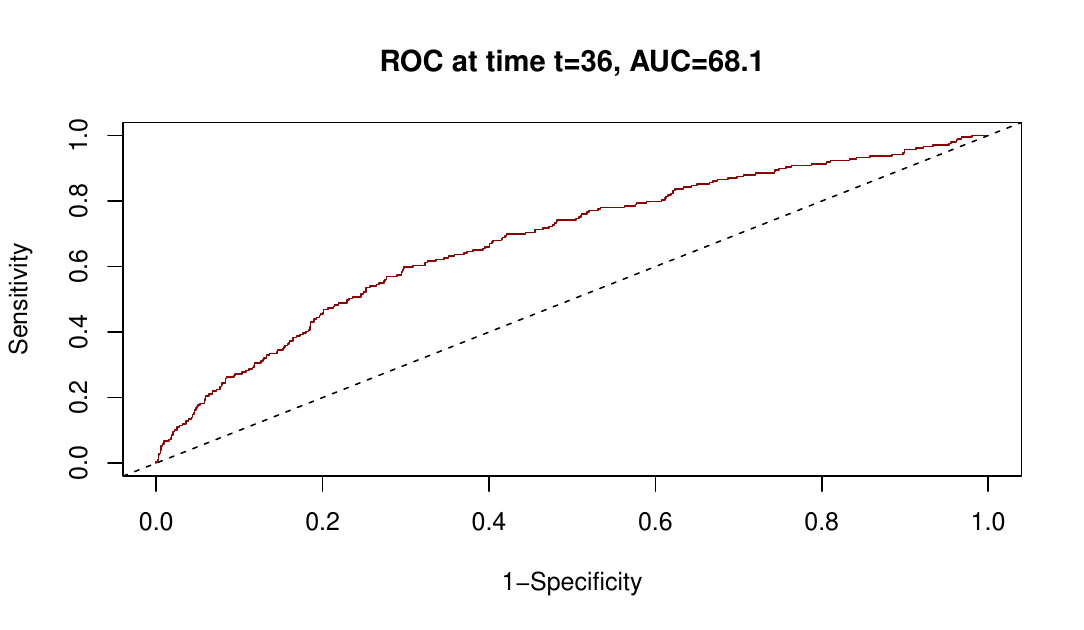}
\caption{Time-dependent AUC curves showing the performance of the reduce Cox model over 36 months. The model achieved an AUC of 0.681 at 36 months, indicating slight discriminative ability.}
\label{roc_final}
\end{figure}

These sensitivity analyses reinforce the robustness of integrating key clinical features with multi-gene  signatures for survival prediction, while also highlighting options for simpler, clinically practical prognostic models.

\noindent
Table~\ref{tab:model_perf} summarizes the comparative performance of different prognostic models. 
The clinical-only Cox model provided modest discrimination (C-index = 0.62, AUC = 0.642), serving as a baseline.  
In contrast, the hybrid Cox model that integrated both clinical and gene expression predictors markedly improved performance (C-index = 0.922, AUC = 0.949), highlighting the benefit of combining molecular and clinical data for risk stratification. 
Finally, the Random Survival Forest (RSF) achieved an AUC of 0.88, demonstrating its ability to capture nonlinear effects, although it did not outperform the integrative Cox model.

\begin{table}[h!]
\centering
\caption{Summary of Model Performance Metrics}
\begin{tabular}{|l|c|c|c|}
\hline
\textbf{Model} & \textbf{C-index} & \textbf{AUC (36 months)} & \textbf{Remarks} \\
\hline
clinical-only      & 0.64 & 0.62 & Baseline \\
Reduce Cox Model      & 0.610 & 0.681 &  \\
Elastic Net (Clin + Gene) & 0.922 & 0.949 & Best performance \\
RSF (OOB estimate)       & ---   & 0.880 & Captures nonlinearity \\
\hline
\end{tabular}
\label{tab:model_perf}
\end{table}

\subsection{Visualization of Gene Expression Patterns}
To explore patterns in gene expression among risk groups, we conducted Principal Component Analysis (PCA) on the standardized expression matrix of the top 20 predictors selected by the Elastic Net model. The first two principal components (Figure) explained approximately 15.3\% of the total variance. When stratified by risk group, patients showed distinguishable clustering in the PCA plot. Confidence ellipses drawn around each group further highlighted the separation, supporting the biological relevance of the risk-based classification.

\begin{figure}[H]
\centering
\includegraphics[width=0.7\textwidth]{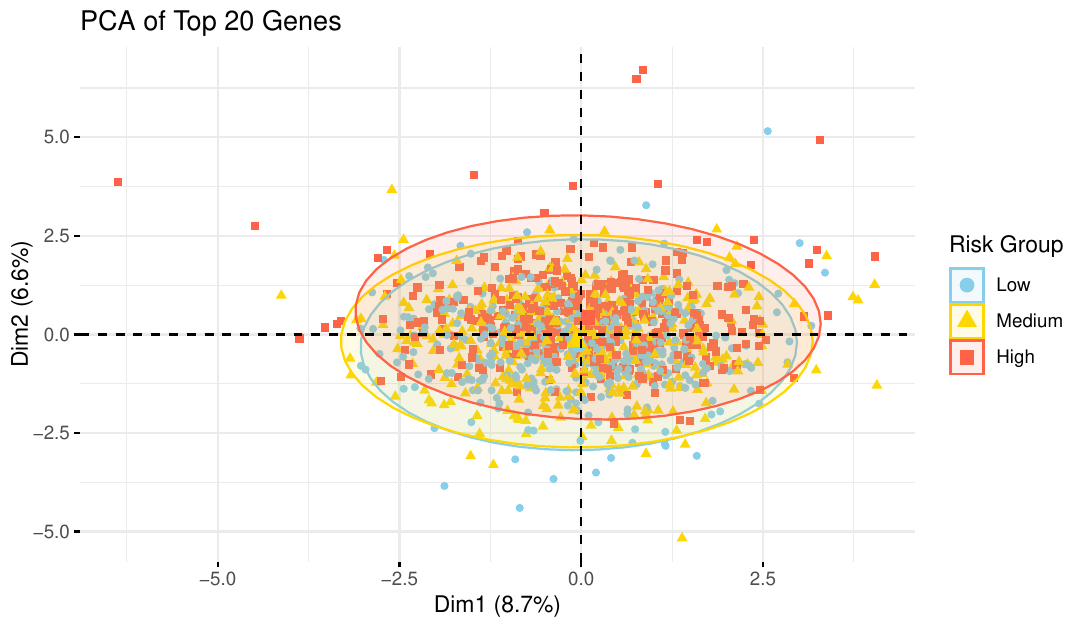}
\caption{Principal component analysis (PCA) plot of selected gene expression and clinical variable. Patients show distinct clustering patterns by model-based risk groups, supporting underlying molecular differentiation.}
\label{PCA_plot}
\end{figure}

In addition, a heat map was generated to visualize the expression patterns of the top 20 genes in all patients. Expression levels were z-score standardized for each gene, and hierarchical clustering was performed using Euclidean distance and complete linkage. Patients were ordered by predicted risk scores from the Cox model, and distinct expression profiles emerged across low-, medium-, and high-risk groups. The heatmap revealed gene clusters with high or low expression concordant with predicted prognosis.

\begin{figure}[H]
\centering
\includegraphics[width=0.8\textwidth]{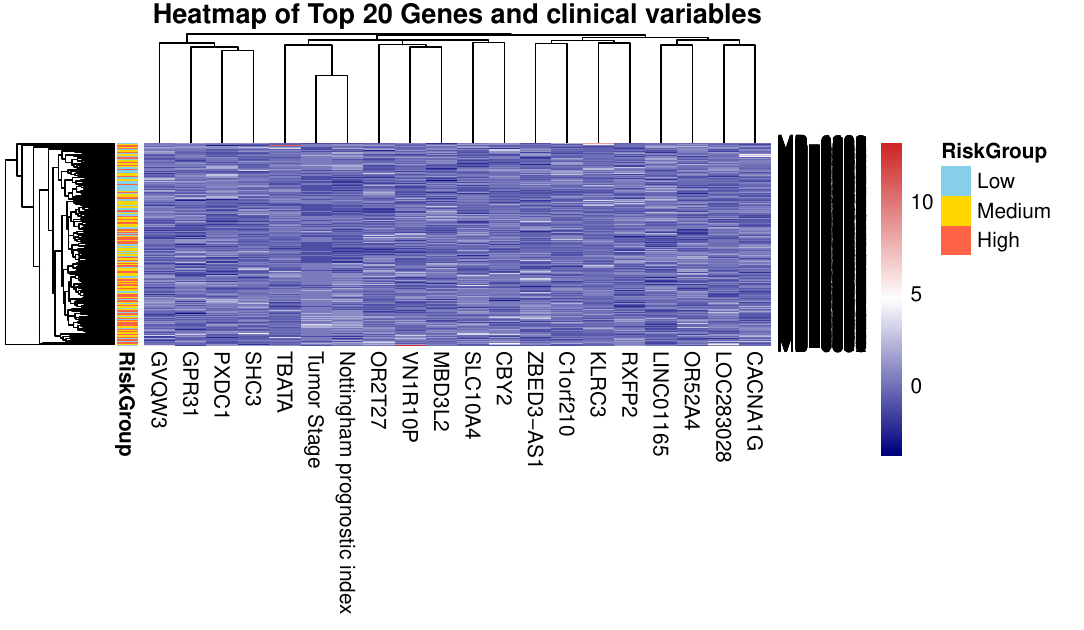}
\caption{Heatmap of expression levels of 20 genes and clinical variable selected by Elastic Net across patient samples. Expression profiles vary clearly across risk groups, suggesting molecular stratification.}
\label{Heatmap_plot}
\end{figure}

 Moreover, Kaplan-Meier and ROC curves for all models (Elastic Net, RSF, and clinical-gene Cox) consistently demonstrated stratified survival, confirming the robustness of our integrative framework.

These visualizations provide an interpretable, hypothesis-generating view of how molecular profiles contribute to survival risk, reinforcing findings from the RSF and Cox models.

\section{Discussion}
This study demonstrates the utility of integrating gene expression signatures with clinical variables to enhance survival prediction in breast cancer patients. By combining Elastic Net penalized Cox regression and Random Survival Forests (RSF), we captured both linear and nonlinear associations between molecular features and clinical outcomes. The final integrative model, comprising genes such as \textit{OR2T27}, \textit{TBATA}, and \textit{LINC01165}, alongside clinical variables like lymph node status, tumor size, and type of surgery, achieved a high concordance index (C-index = 0.922) and 36-month AUC of 0.94, outperforming models using clinical (C = 0.64) or gene-only (C = 0.60) data.

These findings align with recent studies emphasizing the prognostic value of hybrid models.  \cite{Sammut2022} developed a similar model that combined gene expression and clinical variables, reporting improved accuracy over traditional models. Likewise, \cite{Wang2024} demonstrated the advantage of integrating molecular data into machine learning pipelines for survival prediction. Our work extends these findings by validating a multi-step hybrid pipeline using both statistical and machine learning approaches in a large, well-curated cohort.

The Elastic Net was particularly effective in handling high-dimensional gene expression data, enabling sparse and stable variable selection while mitigating multicollinearity \cite{Zou2005}. RSF provided a complementary nonlinear modeling approach, ranking variables by importance and capturing complex interactions that might be overlooked in conventional Cox regression. The RSF model’s out-of-bag (OOB) error rate of 33.7\% and standardized CRPS of 0.163 further confirm the robustness of our integrative approach. This dual framework balances interpretability with flexibility, supporting its potential clinical translation.

Interestingly, several of the top predictors have emerging links to immune signaling and receptor-mediated pathways. For example, \textit{OR2T27} belongs to the olfactory receptor family, which has been implicated in tumor cell proliferation and migration, while \textit{TBATA} and \textit{LINC01165} are associated with immune regulation and transcriptional control. Their consistent selection in our models suggests they may represent underexplored contributors to breast cancer heterogeneity

Visualization techniques, including PCA and heatmaps, provided an interpretable representation of gene expression patterns and risk group clustering. The distinct separation of low-, medium-, and high-risk groups in PCA space and the consistent expression trends observed in heatmaps underscore the biological relevance of the identified gene signatures.

\subsection*{Limitations}
While the integrative models demonstrated strong predictive performance, several limitations must be acknowledged. First, overfitting remains a potential concern, particularly with high-dimensional gene expression data and complex models such as RSF. Although cross-validation was employed, external validation on independent cohorts would strengthen the generalizability of our findings. Second, differences in gene expression measurement platforms (e.g., microarray vs. RNA-seq) may affect model reproducibility across studies. Lastly, our analysis does not account for treatment heterogeneity, which could influence survival outcomes.

\section{Conclusion}
This study presents an integrative prognostic model for breast cancer survival that combines gene expression signatures with key clinical variables. Leveraging Elastic Net penalized Cox regression and Random Survival Forests (RSF), we identified a parsimonious yet highly predictive set of molecular and clinical features. The final model demonstrated strong prognostic performance, achieving a concordance index (C-index) of 0.922, and significantly outperformed models based solely on clinical or gene expression data.

Our findings highlight the added value of multi-omics integration for enhancing prognostic accuracy in breast cancer. Notably, genes such as \textit{OR2T27}, \textit{TBATA}, and \textit{LINC01165}, along with traditional clinical factors like lymph node status and tumor size, emerged as key predictors of survival. These markers hold potential for informing risk stratification and guiding personalized treatment decisions.

Overall, this work underscores the promise of combining high-dimensional molecular data with advanced statistical and machine learning methods for survival modeling. With additional validation in independent cohorts, the proposed framework may contribute to clinical decision-making and advance precision oncology efforts in breast cancer care.

\bibliographystyle{plainnat} 
\bibliography{references}    

\end{document}